\documentclass{article}
\usepackage{eurosym}
\usepackage{amssymb}
\usepackage{amsmath}
\usepackage{cite}

\input{tcilatex}
\begin{document}

\title{Lie symmetries and singularity analysis for generalized shallow-water
equations}
\author{Andronikos Paliathanasis\thanks{%
Email: anpaliat@phys.uoa.gr} \\
{\ \textit{Institute of Systems Science, Durban University of Technology }}\\
{\ \textit{PO Box 1334, Durban 4000, Republic of South Africa}}}
\maketitle

\begin{abstract}
We perform a complete study by using the theory of invariant point
transformations and the singularity analysis for the generalized
Camassa-Holm equation and the generalized Benjamin-Bono-Mahoney equation.
From the Lie theory we find that the two equations are invariant under the
same three-dimensional Lie algebra which is the same Lie algebra admitted by
the Camassa-Holm equation. We determine the one-dimensional optimal system
for the admitted Lie symmetries and we perform a complete classification of
the similarity solutions for the two equations of our study. The reduced
equations are studied by using the point symmetries or the singularity
analysis. Finally, the singularity analysis is directly applied on the
partial differential equations from where we infer that the generalized
equations of our study pass the singularity test and are integrable.

\end{abstract}

\section{Introduction}

The Lie symmetry analysis plays a significant role in the study of nonlinear
differential equations. The existence of a Lie symmetry for a given
differential equation is equivalent with the existence of one-parameter
point transformation which leaves the differential equation invariant. The
later property can be used to reduce the number of independent variables on
the case of partial differential equations (PDE), or reduce the order of an
ordinary differential equation (ODE) \cite{bluman}, that is achieved thought
the Lie invariants. In addition, Lie symmetries can been used for the
determination of conservation laws. One the most well-know applications of
the latter are the two theorems of E. Noether \cite{bluman}. However, there
are also alternative methods to determine the conservation laws by using the
Lie point symmetries without imposing a Lagrange function, some of these
alternative approaches are described in \cite{bluman,ref2,ref3,ref4,ref5}
and references therein.

The are many applications of the Lie symmetries on the analysis of
differential equations, for the determination of exact solutions, to
determine conservation laws, study the integrability of dynamical systems or
classify algebraic equivalent systems \cite%
{rep1,rep2,rep3,rep4,rep5,rep6,rep6a,rep6b}. Integrability is a very
important property of dynamical systems, hence it worth to investigate if a
given dynamical system is integrable \cite%
{rep7,rep8,rep9,rep10,rep11,rep12,rep13,rep14}.

An alternative approach for the study of the integrability of nonlinear
differential equations is the singularity analysis. In contrary with the
symmetry analysis, singularity analysis is based on the existence of a pole
for the differential equation. The first major result of the singularity
analysis is the determination of the third integrable case of Euler's
equations for a spinning by Kowalevskaya \cite{Kowalevski88}. Since then,
the have been many contributions of the singularity analysis, mainly by the
French school led by Painlev\'{e} \cite{pp1,pp2,pp3} and many others \cite%
{b1,b2,b3,b4,b5}. Nowadays, the application of singularity analysis is
summarized in the ARS algorithm \cite{ars1,ars2,ars3} which has made the
singularity analysis a routine tool for the practising applied
mathematicians.

Singularity analysis and symmetry analysis have been applied in a wide range
of differential equations arising from all areas of applied mathematics, for
instance see \cite{c01,c02,c04,c06,c07,c08,c10,c11,c12,c13,c14,c15,c16} and
references therein.\ The two methods are supplementary, on the study of
integrability of differential equations. Usually, the symmetry method is
applied to reduce the given differential equation into an algebraic
equation, or into another well-known integrable differential equation. On
the other hand, for differential equation which posses the Painlev\'{e}
property, i.e. it pass the singularity test, its solution is written in
terms of Laurent expansions, a recent comparison of the two methods is
presented in \cite{leachan}.

In this work, we study the integrability of generalized Camassa-Holm (CH)
and Benjamin-Bono-Mahoney (BBM) equations \cite{ch1,ch2} by using the Lie
point symmetries and the singularity analysis. These two equations describe
shallow-water phenomena.

The Camassa-Holm equation is a well-known integrable equation. It was
originally discovered by Fuchssteiner et al. in \cite{focas}, however become
popular a decade later by the study of Camassa and Holm where they proved
the existence of peaked solutions, also known as peakons. On the other hand,
BBM equation also known as regularized long-wave equation discovered in \cite%
{bb01} and it is an extension of the KdV equation. The two equations are
related, in the sense they have a common operator and partial common
Hamiltonian structure. The plan of the paper is as follows.

In Section \ref{sec2} we present the basic elements on the mathematical
tools of our consideration, that is, the Lie point symmetries and the
singularity analysis. Our main analysis is included in Sections \ref{sec3}
and \ref{sec4} where we study the existence of similarity solutions for the
generalized CH and BBM equations, as also we prove the integrability of
these two equations by using the singularity analysis. Finally, we discuss
our results and draw our conclusions in Section \ref{sec5}.

\section{Preliminaries}

\label{sec2}

In this section we briefly discuss the application of Lie's theory on
differential equations as also the main steps of the singularity analysis.

\subsection{Lie symmetries}

Consider the vector field 
\begin{equation}
X=\xi ^{i}\left( x^{k},u\right) \partial _{i}+\eta \left( x^{k},u\right)
\partial _{u},  \label{go.10}
\end{equation}%
to be the generator of the local infinitesimal one-parameter point
transformation, \ 
\begin{align}
\bar{x}^{k}& =x^{k}+\varepsilon \xi ^{i}\left( x^{k},u\right) , \\
\bar{\eta}& =\eta +\varepsilon \eta \left( x^{k},u\right) .
\end{align}

Then $X$ is called a Lie symmetry for the differential equation, $\mathbf{H}%
\left( y^{i},u,u_{i},u_{ij},...,u_{i_{1}i_{2}...i_{n}}\right) $, if there
exists a function $\lambda $ such that the following condition to hold%
\begin{equation}
X^{\left[ n\right] }H=\lambda H  \label{go.11}
\end{equation}%
where $X^{\left[ n\right] }$ is called the second prolongation/extension in
the jet-space and is defined as%
\begin{equation}
X^{\left[ n\right] }=X+\left( D_{i}\eta -u_{,k}D_{i}\xi ^{k}\right) \partial
_{u_{i}}+\left( D_{i}\eta _{j}^{\left[ i\right] }-u_{jk}D_{i}\xi ^{k}\right)
\partial _{u_{ij}}+...+\left( D_{i}\eta _{i_{1}i_{2}...i_{n-1}}^{\left[ i%
\right] }-u_{i_{1}i_{2}...k}D_{i_{n}}\xi ^{k}\right) \partial
_{u_{i_{1}i_{2}...i_{n}}}.  \label{go.13}
\end{equation}

The novelty of Lie symmetries is that they can be used to determine
similarity transformations, i.e. differential transformations where the
number of independent variables is reduced \cite{bluman}. The similarity
transformation is calculated with the use of the associated Lagrange's
system, 
\begin{equation}
\frac{dx^{i}}{\xi ^{i}}=\frac{du}{u}=\frac{du_{i}}{u_{\left[ i\right] }}=...=%
\frac{du_{ij..i_{n}}}{u_{\left[ ij...i_{n}\right] }}.
\end{equation}

The similarity transformation in the case of PDEs is used to reduce the
number of indepedent variables. The solutions derived by the application of
Lie invariants are called similarity solutions.

\subsection{Singularity analysis}

The modern treatment of the singularity analysis is summarized in the ARS
algorithm, established by Ablowitz, Ramani and Segur in \cite{ars1,ars2,ars3}%
. There are three basic steps which are summarized as follows: (a) determine
the leading-order term which describes the behaviour of the solution near
the singularity, (b) find the position of the resonances \ which shows the
existence and the position of the integration constants and (c) write a
Laurent expansion with leading-order term determined in step (a) and perform
the consistency test. More details on the ARS algorithm as also on the
conditions which should hold at every step we refer the reader in n the
review of Ramani et al. \cite{buntis}, where illustrated applications are
presented.

It is important to mention that when a differential equation passes the
conditions and requirement of the ARS algorithm we can infer that the given
differential equation is algebraically integrable.

\section{Generalized Camassa-Holm equation}

\label{sec3}

We work with the generalized CH equation defined in \cite{ch1,ch2} 
\begin{equation}
u_{t}-u_{xxt}+\frac{\left( k+2\right) \left( k+1\right) }{2}%
u^{k}u_{x}=\left( \frac{k}{2}u^{k-1}u_{x}^{2}+u^{k}u_{xx}\right) _{x}~,~
\label{ch.01}
\end{equation}%
where $k\geq 1$ is a positive integer number, while when $k=1$ CH equation
is recovered. The Lie symmetry analysis for the CH equation presented before
in \cite{c04}. It was found that the CH is invariant under a three
dimensional Lie algebra.

For the generalized CH equation (\ref{ch.01}) the application of Lie's
theory provides us that the admitted Lie point symmetries are three, and
more specifically they are%
\begin{equation*}
X_{1}=\partial _{t}~,~X_{2}=\partial _{x}~\text{and }X_{3}=t\partial
_{t}-ku\partial _{u}.
\end{equation*}%
The commutators and the adjoint representation of the Lie point symmetries
are presented in Tables \ref{tab1} and \ref{tab2} respectively.

The results presented in Tables \ref{tab1} and \ref{tab2} can be used to
classify the admitted Lie algebra as also to determine the one-dimensional
optimal systems \cite{olverb}. A necessary analysis to perform a complete
classification of the similarity solutions.

As far as the admitted Lie algebra is concerned, from Table \ref{tab1} it is
find to be the $\left\{ 2A_{1}\otimes _{s}A_{1}\right\} $ in the
Morozov-Mubarakzyanov Classification Scheme \cite%
{Morozov58a,Mubarakzyanov63a,Mubarakzyanov63b,Mubarakzyanov63c}.

In order to find the one-dimensional optimal systems we consider the generic
symmetry vector 
\begin{equation}
\mathbf{X}=a_{1}X_{1}+a_{2}X_{2}+a_{3}X_{3},  \label{ad.02}
\end{equation}%
from where we find the equivalent symmetry by considering the adjoint
representation. We remark that the adjoint action admits two invariant
functions the $\phi _{1}\left( a_{i}\right) =a_{3}$ and $\phi _{2}\left(
a_{i}\right) =a_{3}$ which are necessary to simplify the calculations on the
derivation of the one-dimensional systems. More specifically there are four
possible cases, $\left\{ \phi _{1}\phi _{2}\neq 0\right\} $,$~\left\{ \phi
_{1}\neq 0~,~\phi _{2}=0\right\} ~,~\left\{ \phi _{1}=0~,~\phi _{2}\neq
0\right\} $ and $\left\{ \phi _{1}=0~,~\phi _{2}=0\right\} $.

Consequently, with the use of the invariant functions $\phi _{1}$ and $\phi
_{2}$ and Table \ref{tab2} we find that the possible one-dimensional optimal
systems are%
\begin{equation}
X_{1},~X_{2}~,~X_{3}~,~cX_{1}+X_{2}~\text{and }X_{2}+\alpha C_{3}.
\label{ad.03}
\end{equation}%
We proceed with the application of the latter one-dimensional system in
order to reduce the PDE (\ref{ch.01}) into an ODE.

\begin{table}[tbp] \centering%
\caption{Commutators of the admitted Lie point symmetries by the
differential equation (\ref{ch.01})}%
\begin{tabular}{cccc}
\hline\hline
$\left[ ,\right] $ & $\mathbf{X}_{1}$ & $\mathbf{X}_{2}$ & $\mathbf{X}_{3}$
\\ \hline
$\mathbf{X}_{1}$ & $0$ & $0$ & $kX_{1}$ \\ 
$\mathbf{X}_{2}$ & $0$ & $0$ & $0$ \\ 
$\mathbf{X}_{3}$ & $-kX_{1}$ & $0$ & $0$ \\ \hline\hline
\end{tabular}%
\label{tab1}%
\end{table}%

\begin{table}[tbp] \centering%
\caption{Adjoint representation for the Lie point symmetries of the
differential equation (\ref{ch.01})}%
\begin{tabular}{cccc}
\hline\hline
$Ad\left( \exp \left( \varepsilon X_{i}\right) \right) X_{j}$ & $\mathbf{X}%
_{1}$ & $\mathbf{X}_{2}$ & $\mathbf{X}_{3}$ \\ \hline
$\mathbf{X}_{1}$ & $X_{1}$ & $X_{2}$ & $X_{3}-k\varepsilon X_{1}$ \\ 
$\mathbf{X}_{2}$ & $X_{1}$ & $X_{2}$ & $X_{1}$ \\ 
$\mathbf{X}_{3}$ & $e^{\varepsilon k}X_{1}$ & $X_{2}$ & $X_{3}$ \\ 
\hline\hline
\end{tabular}%
\label{tab2}%
\end{table}%

\subsection{Analytic solutions}

In this section we proceed with the application of the\ Lie point symmetries
to the nonlinear generalized CH equation. In order to solve the reduced
equation we apply the Lie point symmetries and when it is no possible to
proceed the reduction process, we consider the singularity analysis by
applying the ARS algorithm.

\subsubsection{Reduction with $X_{1}:$ Static solution}

The application of the Lie point symmetry vector $X_{1}$ indicates that the
solution $u$ is static, i.e. $u=U\left( x\right) $ where now function $%
U\left( x\right) $ satisfies the third-order ODE%
\begin{equation}
\frac{\left( k+2\right) \left( k+1\right) }{2}U^{k}U_{x}-\left( \frac{k}{2}%
U^{k-1}U_{x}^{2}+U^{k}U_{xx}\right) _{x}=0~.  \label{ch.04}
\end{equation}%
The latter equation can be easily integrated, and be written in the
equivalent form%
\begin{equation}
U_{xx}+\frac{k}{2U}U_{x}^{2}-\frac{U_{0}}{U^{k}}+\frac{\left( k+2\right) }{2}%
U=0,  \label{ch.05}
\end{equation}%
where $U_{0}$ is a constant of integration. Equation (\ref{ch.05}) admits
the following conservation law%
\begin{equation}
\frac{1}{2}U^{k}\left( U_{x}\right) ^{2}-U_{0}U-\frac{U^{K+2}}{2}=U_{1},
\label{ch.06}
\end{equation}%
in which $U_{1}$ is a second constant of integration. Equation (\ref{ch.06})
can be integrated by quadratures.

Conservation law (\ref{ch.06}) is nothing else than the Hamiltonian function
of the second-order ODE (\ref{ch.05}). Before we proceed with another
reduction let us now apply the singularity analysis to determine the
analytic solution of equation (\ref{ch.04}).

\paragraph{Singularity analysis}

We substitute $U\left( x\right) =U_{0}\chi ^{p},~\chi =x-x_{0},$ in (\ref%
{ch.04}) and we find the polynomial expression%
\begin{equation}
\left( k+1\right) \left( k+2\right) \chi ^{p\left( k+1\right) -1}+\left(
p\left( k+1\right) -2\right) \left( p\left( k+2\right) -2\right) \chi
^{p\left( k+1\right) -3}=0,  \label{ch.07}
\end{equation}%
Hence we can infer that the only possible leading-order behaviour to the
terms with $x^{p\left( k+1\right) -3}$, where from the requirement 
\begin{equation}
\left( p\left( k+1\right) -2\right) \left( p\left( k+2\right) -2\right) =0,
\label{ch.08}
\end{equation}%
provides 
\begin{equation}
p_{1}=\frac{2}{k+2}~\text{or }p_{2}=\frac{2}{k+1},  \label{ch.09}
\end{equation}%
while constant $U_{0}$ is undetermined.

Consider now the leading order term $p_{1}$. In order to find the resonances
we replace $U\left( x\right) =U_{0}\chi ^{p_{1}}+\mu \chi ^{p_{1}+s}$ in (%
\ref{ch.09}) and we linearize around the $\mu =0$. From the linear terms of $%
\mu $, the coefficient of the leading order terms $\chi ^{-\frac{2}{2+k}+s}$
are $s\left( s+1\right) \left( s\left( k+2\right) -2\right) $, where the
requirement the latter expression to be zero provides the three resonances%
\begin{equation}
s_{1}=-1~,~s_{2}=0\text{ and }s_{3}=\frac{2}{k+2}.
\end{equation}%
We remark that because $k$ is always a positive integer number, $p_{1}$ and $%
s_{3}$ are always rational numbers. Resonance $s_{1}$ indicates that the
singularity is movable, the position of the singularity is one of the three
integration constants. Resonances $s_{2}$ shows that the coefficient
constant of the leading-order term should be arbitrary since it is also one
of the integration constants of the problem. The third constant it is given
at the position of the resonance $s_{3}$ and depends on the value $k$.
Moreover, because all the resonances are positive the solution will be given
by a Right Painlev\'{e} Series. In order to complete the ARS algorithm we
should perform the consistency test. For that we select a special value of $%
k $.

We select $k=2$, and we consider the right Laurent expansion%
\begin{equation}
U\left( x\right) =U_{0}\chi ^{\frac{1}{2}}+U_{1}\chi +U_{2}\chi ^{\frac{3}{2}%
}+\dsum\limits_{i=3}^{\infty }U_{i}\chi ^{\frac{1}{2}+\frac{i}{2}},
\label{ch.10}
\end{equation}%
we find that $U_{1}$ is the third integration constant while the first
coefficient constants are 
\begin{equation}
U_{2}\left( U_{0},U_{1}\right) =-\frac{7}{8}\frac{\left( U_{1}\right) ^{2}}{%
U_{0}},~U_{3}=\frac{5}{4}\frac{\left( U_{1}\right) ^{3}}{\left( U_{0}\right)
^{2}}~,~U_{4}=-\frac{273}{128}\frac{\left( U_{1}\right) ^{4}}{\left(
U_{0}\right) ^{3}}+\frac{U_{0}}{3}~,~...;
\end{equation}%
Hence, the consistency test is satisfied and expression (\ref{ch.10}) is one
solution of the third-order ODE (\ref{ch.04}).

We work similar and for the second leading-order behaviour $p_{2}=\frac{2}{%
k+1}$, the resonances are derived to be%
\begin{equation}
s_{1}=-1~,~s_{2}=0~,~s_{3}=-\frac{2}{k+1},
\end{equation}%
from where we infer that the solution is given by a Mixed Painlev\'{e}
Series. However, we perform the consistency test for various values of the
integer number $k$, and we conclude that for this leading-order behaviour
the differential equation does not pass the singularity test.

In order to understand better why only the leading-order behaviour $p_{1}$
passes the singularity test, let us perform the same analysis for the
second-order ODE (\ref{ch.05}). By replacing in $U\left( x\right) =U_{0}\chi
^{p}$ in (\ref{ch.05}) we find that the unique leading-order behaviour is
that with $p=\frac{2}{k+2}$ with arbitrary $U_{0}.$ The two resonances now
are calculated to be $s_{1}=-1$ and $s_{2}=0$, from where we can infer that
the solution is given by a Right Painlev\'{e} series and the two integration
constants is the $U_{0}$ and the position of the singularity. In that case,
since we know the two integration constants it is not necessary to perform
the consistency test.

\subsubsection{Reduction with $X_{2}$: Stationary solution}

Reduction with the vector field $X_{2}$ provides the stationary solution $%
u\left( t,x\right) =U\left( t\right) $, where $U_{t}=0$, that is $u\left(
t,x\right) =u_{0}$. This is the trivial solution.

\subsubsection{Reduction with $X_{3}$: Scaling solution I}

From the symmetry vector $X_{3}$ we derive the Lie invariants%
\begin{equation}
u=U\left( x\right) t^{-\frac{1}{k}}~,~x  \label{ch.12}
\end{equation}%
hence, by replacing in (\ref{ch.01}) we end up with the following
third-order ODE%
\begin{equation}
2kU^{k}U_{xxx}-2\left( 1-2k^{2}U^{k-1}U_{x}\right) U_{xx}+k\left( \left(
1-k\right) kU^{k-2}\left( U_{x}\right) ^{2}-\left( k+2\right) \left(
k+1\right) U^{k}\right) U_{x}+2U=0,  \label{ch.13}
\end{equation}%
The latter equation admits only one Lie point symmetry, the vector field $%
X_{2}.$ The latter vector field can be used to reduce equation (\ref{ch.13})
into a second-order nonautonomous ODE, with no symmetries. Hence, we proceed
with the application of the singularity analysis for equation (\ref{ch.13}).

We replace $U\left( x\right) =U_{0}\chi ^{p}$ in (\ref{ch.13}) where we find
the following expression 
\begin{equation}
-2U_{0}\chi ^{p}+2U_{0}p\left( p-1\right) \chi ^{p-2}+U_{0}^{k+1}k\left(
k+2\right) \left( k+1\right) \chi ^{p\left( k+1\right)
-1}-U_{0}^{k+1}pk\left( p\left( k+2\right) -2\right) \left( p\left(
k+1\right) -2\right) \chi ^{\left( p\left( k+1\right) -3\right) }=0.
\end{equation}%
From the latter term we find that the only possible leading terms with $k$
positive integer number are $p-2=\left( p\left( k+1\right) -3\right) ~$from
where we find that $p=\frac{1}{k}$ while $U_{0}$ is given by the following
expression%
\begin{equation}
U_{0}^{-k}=2\left( 2-k\right) ,
\end{equation}%
from where we infer that there is a leading-order behaviour only for $k\neq
2 $.

The resonances are calculated to be 
\begin{equation}
s_{1}=-1~,~s_{2}=\frac{k-1}{k}~,~s_{2}=\frac{k-2}{2k},
\end{equation}%
from where we can infer that for $k>2$, the solution is given by a Right
Painlev\'{e} Series. We perform the consistency test by choosing $k=3$.
Hence the Laurent expansion is written as 
\begin{equation}
U\left( \chi \right) =U_{0}x^{\frac{1}{3}}+U_{1}x^{\frac{1}{2}%
}+\dsum\limits_{i=2}^{\infty }U_{i}\chi ^{\frac{1+2i}{3}},
\end{equation}%
where $U_{1}$ and $U_{4}$ are two integration constants of the solution
while the rest coefficient constants are $U_{i}=U_{i}\left(
U_{1},U_{4}\right) $.

\subsubsection{Reduction with $cX_{1}+X_{2}$: Travel-wave solution}

The travel-wave similarity solution is determined by the application of the
Lie invariants of the symmetry vector $cX_{1}+X_{2}$ where $c^{-1}$ is the
travel-wave speed. The invariant functions for that vector field are
determined to be%
\begin{equation}
u\left( t,x\right) =U\left( \xi \right) ~,~\xi =x-c^{-1}t,
\end{equation}%
where $U\left( \xi \right) $ satisfies the following third-order ODE%
\begin{equation}
2\left( 1-cU^{k}\right) U_{\xi \xi \xi }-4ckU^{k-1}U_{\xi }U_{\xi \xi
}-\left( c\left( k-1\right) U^{k-1}U_{\xi \xi }+2-c\left( k+2\right) \left(
k+1\right) U^{k}\right) U_{\xi }=0.  \label{ch.14a}
\end{equation}

Equation (\ref{ch.14a}) is autonomous and admit only one symmetry vector the 
$\partial _{\xi }$. It can easily integrated as follows%
\begin{equation}
2\left( 2-cU^{k}\right) U_{\xi \xi }-c\left( 2U^{k-2}-U^{k-1}\right) \left(
U_{\xi }\right) ^{2}+c\left( k+2\right) U^{k-3}-2U+U_{0}=0,  \label{ch.14}
\end{equation}%
which can be solved by quadratures.

Let us now apply the singularity analysis to write the analytic solution of
equation in (\ref{ch.14}) by using Laurent expansions. We apply the ARS
algorithm and we find the leading order term $U\left( \xi \right)
=U_{0}\left( \xi -\xi _{0}\right) ^{p}$ with $p=\frac{2}{k+2}$ and $U_{0}$
arbitrary. The resonances are calculated to be $s_{1}=0$ and $s_{2}=0$,
which means that the solution is given by a Right Painlev\'{e} Series with
integration constants the position of the singularity $\xi _{0}$ and the
coefficient constant of the leading order term $U_{0}$. The step of the
Painlev\'{e} Series \ depends on the value of $k$, for instance for $k=2$, \ 
$p=\frac{1}{2}$ and the step is $\frac{1}{2}$, while for $k=3,~p=\frac{2}{5}$
and the step is $\frac{1}{5}$.

\subsubsection{Reduction with $X_{2}+\protect\alpha X_{3}$: Scaling solution
II}

We complete our analysis by determine the similarity solution given by the
symmetry vector $X_{2}+\alpha X_{3}$. The that specific symmetry the Lie
invariants are calculated%
\begin{equation}
u\left( t,x\right) =U\left( \xi \right) t^{-\frac{1}{k}}~,~\xi =x+\frac{1}{%
\alpha k}\ln t.
\end{equation}

Therefore, by selecting $\xi $ to be the new independent variable and $%
U\left( \xi \right) $ the new dependent variable we end up with the
third-order ODE 
\begin{equation}
2\left( 1+\alpha k\right) U_{\xi \xi \xi }-2\alpha \left(
1-2k^{2}U^{k-1}U_{\xi }\right) U_{\xi \xi }+\left( a\left( k-1\right)
U^{k-2}\left( U_{\xi }\right) ^{2}-2-\alpha kU^{k}k\left( k+2\right) \left(
k+1\right) \right) U_{\xi }+2\alpha U=0.  \label{ch.16}
\end{equation}

The latter equation is autonomous and admit only one point symmetry, the
vector field $\partial _{\xi }$, which can be used to reduce by one the
order of the ODE. The resulting second-order ODE has no symmetries. Hence,
the singularity analysis is applied to study the integrability of (\ref%
{ch.16}).

In order to perform the singularity analysis we do the change of variable $%
V=U^{-1}$. Hence by replacing $V\left( \xi \right) =V_{0}\left( \xi -\xi
_{0}\right) ^{p}$ in (\ref{ch.16}) we find the leading-order terms 
\begin{equation}
p_{1}=-1\text{ and }p_{2}=-2~~\text{for }k>1,
\end{equation}%
while $V_{0}$ is arbitrary.

The resonances are calculated to be

\begin{eqnarray}
p_{1} &:&s_{1}=-1~,~s_{2}=0\text{ and }s_{3}=1;  \label{ch.17} \\
p_{2} &:&s_{1}=-1~,~s_{2}=0~\text{and}~s_{3}=-2.  \label{ch.18}
\end{eqnarray}%
We apply the consistency test where we find that only only the leading-order
term $p_{1}$ provides a solution, which is given by the following Right
Painlev\'{e} Series%
\begin{equation*}
V\left( \xi \right) =V_{0}\left( \xi -\xi _{0}\right)
^{-1}+\dsum\limits_{i=1}^{\infty }V_{i}\left( \xi -\xi _{0}\right) ^{-1+i}.
\end{equation*}

\subsection{Singularity analysis}

\label{sinch}

Until now we applied the singularity analysis to study the integrability of
the ODEs which follow by the similarity reduction for the generalized CH
equation. However, it is possible to apply the singularity analysis directly
in the PDE. We follow the steps presented in \cite{b5}.

Before we proceed with the application of the ARS algorithm we make the
change of transformation $u\left( t,x\right) =v\left( t,x\right) ^{-1}$ in (%
\ref{ch.01}). For the new variable we search for a singular behaviour of the
form $v\left( t,x\right) =v_{0}\left( t,x\right) \phi \left( t,x\right) ^{p}$%
, where $v_{0}\left( t,x\right) $ is the coefficient function and $\phi
\left( t,x\right) ^{p}$ is the leading-order term which describe the
singularity.

The first step of the ARS algorithm provides two values of $p,$ $p_{1}=-1$
and~$p_{2}=-2,~$\ where $v_{0}\left( t,x\right) $ is arbitrary. A necessary
and sufficient condition in order these two leading-order terms to exists is 
$\phi _{,t}\phi _{,x}\neq 0$. Otherwise other leading-order terms follow,
however these possible cases studied before. The resonances for these two
leading-order terms are those given in (\ref{ch.17}) and (\ref{ch.19}).

Consequently, the following two Painlev\'{e} Series should be studied for
the consistency test%
\begin{equation}
v\left( t,x\right) =v_{0}\left( t,x\right) \phi \left( t,x\right)
^{-1}+\dsum\limits_{i=1}^{\infty }v_{i}\left( t,x\right) \phi \left(
t,x\right) ^{-1+i},  \label{ch.19}
\end{equation}%
\begin{equation}
v\left( t,x\right) =v_{0}\left( t,x\right) \phi \left( t,x\right)
^{-2}+\dsum\limits_{i=1}^{\infty }v_{i}\left( t,x\right) \phi \left(
t,x\right) ^{-2+i}.  \label{ch.20}
\end{equation}

By replacing (\ref{ch.19}) we find that the second integration constant is $%
v_{1}\left( t,x\right) $. On the other hand, the series (\ref{ch.20}) does
not pass the consistency test. We conclude that that the generalized CH
equation passes the singularity test and it is an integrable equation.

We proceed our analysis with the BBM equation.

\section{Generalized Benjamin-Bono-Mahoney equation}

\label{sec4}

The generalized BBM equation is%
\begin{equation}
u_{t}-u_{xxt}+\beta u^{k}u_{x}=0,  \label{bbm.01}
\end{equation}%
where $k$ is a positive integer number. Equation (\ref{bbm.01}) can been
seen as the lhs of (\ref{ch.01}) when $\beta =\frac{\left( k+2\right) \left(
k+1\right) }{2}$ and reduce to the BBM equation when $k=1$. For the case of $%
k=1$ the Lie symmetry analysis for the BBM equation presented recently in 
\cite{bb02a,bb02}.

We apply the Lie theory in order to determine the point transformations
which leave equation (\ref{bbm.01}) invariant. We found that the equation (%
\ref{bbm.01}) admits three point symmetries which are the vector fields $%
X_{1},~X_{2},~X_{3}$ presented in\ Section \ref{sec3}. Hence, the admitted
Lie algebra is the $2A_{1}\otimes _{s}A_{1}$ and there are five
one-dimensional optimal systems as presented in (\ref{ad.03}). We proceed
with the application of the Lie point symmetries for the determination of
similarity solutions.

\subsection{Analytic solutions}

For equation (\ref{bbm.01}) the application of the Lie symmetries $X_{1}$
and $X_{2}$ provide the trivial solution $u\left( t,x\right) =u_{0}$ for
both cases.

\subsubsection{Reduction with $X_{3}$: Scaling solution I}

The application of the Lie invariants which given by the symmetry vector $%
X_{3}$ gives~$u\left( t,x\right) =U\left( x\right) t^{\frac{1}{k}}$ where $%
U\left( x\right) $ satisfies the second-order ODE%
\begin{equation}
U_{xx}+\beta kU^{k}U_{x}-U=0.  \label{bbm.02}
\end{equation}%
The latter equation is autonomous and admit the point symmetry $\partial
_{x} $ which can be used to reduce equation (\ref{bbm.02}) into the
following first-order ODE%
\begin{equation}
y_{z}=\beta kz^{k}y^{2}-zy^{3},
\end{equation}%
where $y\left( z\right) =\left( U_{x}\right) ^{-1}$ and $z=U\left( x\right) $%
.

However, equation (\ref{bbm.02}) can be easily solved analytical by using
the singularity analysis. Indeed from the ARS algorithm we find the
leading-order behaviour%
\begin{equation}
U\left( \chi \right) =\left( \frac{k+1}{\beta k^{2}}\right) ^{\frac{1}{k}%
}x^{-\frac{1}{k}},
\end{equation}%
with resonances 
\begin{equation}
s_{1}=-1\text{ and~}s_{2}=\frac{1+k}{k}.
\end{equation}

In order to perform the consistency test we have to select specific value
for the parameter $k$. Indeed for $k=2$ we write the Laurent expansion%
\begin{equation}
U\left( \chi \right) =\left( \frac{3}{4\beta }\right) ^{\frac{1}{2}}x^{-%
\frac{1}{2}}+\dsum\limits_{i=1}^{\infty }U_{i}x^{-\frac{1+i}{2}},
\end{equation}%
and by replacing in (\ref{bbm.02}) we find that 
\begin{equation}
U_{1}=0~,~U_{2}=0~,~U_{4}=-\sqrt{\frac{1}{3\beta }}~,~U_{5}=0,~U_{6}=-\frac{%
\sqrt{3\beta }}{4}\left( U_{3}\right) ^{2}~,~....
\end{equation}%
where $U_{3}$ is the second integration constant. We conclude that that the
equation (\ref{bbm.02}) passes the Painlev\'{e} test.

\subsubsection{Reduction with $cX_{1}+X_{2}$: Travel-wave solution}

The travel-wave solution of the generalized BBM equation is $u=U\left( \xi
\right) $, where $\xi =x-c^{-1}t$ and $U\left( \xi \right) $ satisfies the
differential equation%
\begin{equation}
U_{\xi \xi \xi }+\left( c\beta U^{k}-1\right) U_{\xi }=0.
\end{equation}%
The latter equation can be integrated easily 
\begin{equation}
U_{\xi \xi }+\left( \frac{c\beta }{k+1}U^{k+1}-U\right) +U_{0}=0,
\label{bhm.01}
\end{equation}%
that is 
\begin{equation}
\frac{1}{2}\left( U_{\xi }\right) ^{2}+\left( \frac{c\beta }{\left(
k+1\right) \left( k+2\right) }U^{k+2}-\frac{U^{2}}{2}\right) +U_{0}U-U_{1}=0,
\end{equation}%
where $U_{0},~U_{1}$ are two integration constants. The latter differential
equation can be solved easily by quadratures.

As far as the singularity analysis is concerned for equation (\ref{bhm.01}),
the ARS algorithm provides the leading-order behaviour 
\begin{equation}
U\left( \xi \right) =U_{0}\left( \xi -\xi _{0}\right) ^{-\frac{2}{k}%
}~,~~U_{0}^{k}=-2\frac{\left( k+1\right) \left( k+2\right) }{\beta ck^{2}},
\end{equation}%
with resonances%
\begin{equation*}
s_{1}=-1\text{ and }s_{2}=\frac{2\left( k+2\right) }{k}.
\end{equation*}%
The consistency test has been applied for various values of the positive
integer $k$, and we can infer that equation (\ref{bhm.01}) is integrable
according to the singularity analysis.

\subsubsection{Reduction with $X_{2}+\protect\alpha X_{3}$: Scaling solution
II}

From the Lie symmetry $X_{2}+\alpha X_{3}$ we find the similarity reduction $%
u\left( t,x\right) =U\left( \xi \right) t^{-\frac{1}{k}}~,~\xi =x+\frac{1}{%
\alpha k}\ln t$ where $U\left( \xi \right) $ is a solution of the following
differential equation%
\begin{equation}
U_{\xi \xi \xi }-\alpha U_{\xi \xi }-\left( \alpha \beta U^{k}+1\right)
U_{\xi }+\alpha U=0.  \label{bbm.03}
\end{equation}
Equation\ (\ref{bbm.03}) can be reduced to the following second-order ODE by
use of the point symmetry vector $\partial _{\xi }$, 
\begin{equation}
z^{2}y_{zz}+z\left( y_{z}\right) ^{2}-\alpha zy_{z}-\left( \alpha \beta
z^{k}+1\right) y+\alpha z=0,  \label{bbm.04}
\end{equation}%
where $z=U\left( x\right) $ and $y\left( z\right) =U_{x}$.

We apply the ARS algorithm for equation (\ref{bbm.03}) and we find that it
passes the singularity test for the leading order behaviour%
\begin{equation}
U\left( \xi \right) =U_{0}\left( \xi -\xi _{0}\right) ^{-\frac{2}{k}%
}~,~U_{0}^{k}=2\frac{\left( k+2\right) \left( k+1\right) }{\alpha \beta k^{2}%
},
\end{equation}%
with resonances%
\begin{equation}
s_{1}=-1~,~s_{2}=\frac{2\left( k+1\right) }{k}~,~s_{3}=\frac{2\left(
k+2\right) }{k}.  \label{bbm.05}
\end{equation}

\subsection{Singularity analysis}

We complete our analysis by applying the singularity test in the generalized
BBM equation in a similar way as we did in Section \ref{sinch} for the
generalized CH equation. Indeed we find the leading order term%
\begin{equation}
u\left( t,x\right) =v_{0}\left( t,x\right) \phi \left( t,x\right) ^{-\frac{2%
}{k}}~,~\left( v_{0}\left( t,x\right) \right) ^{k}=2\frac{\left( k+2\right)
\left( k+1\right) }{\alpha \beta k^{2}}\phi _{t}\phi _{x},  \label{bbm.06}
\end{equation}%
and resonances those given in (\ref{bbm.05}). We performed the consistency
test and we infer that the generalized BBM equation passes the singularity
test for any value of the positive integer parameter $k$.

\section{Conclusion}

\label{sec5}

In this work we studied the existence of similarity solutions of the
generalized CH and generalized\ BBM equation. The approached that we used is
that of the Lie point symmetries. We determined the admitted invariant point
transformations for the differential equations of our consideration and we
determined the one-dimensional optimal systems by using the adjoint
representation of the admitted Lie algebra. The two differential equations
of our consideration are invariant under the same Lie symmetry vectors which
form the same Lie algebra with the CH and the BBM equations.

For each of the equations we perform five different similarity reductions
where the PDEs are reduced to third-order ODEs. The integrability of the
resulting equations is studied by using symmetries and/or the singularity
analysis. In the case of the generalized CH equation most of the reduced
ODEs can not be solved by using Lie symmetries, hence the application of the
singularity analysis was necessary to determine the analytic solutions of
the reduced equations.

Finally, we study the integrability of the PDEs of our consideration by
applying the singularity analysis directly on the PDEs and not on the
reduced equations. From the latter analysis we found that the generalized CH
and BBM equations pass the singularity analysis and their solutions are
given by Right Painlev\'{e} Series.

This work contribute to the subject of the integrability of generalized
equations describe shallow-water phenomena. The physical implication of the
new analytic solutions will be presented in a future communication.

\end{document}